%% file: main.tex
\documentclass[%
 reprint,prl
 amsmath,amssymb,
 aps, physrev,
]{revtex4-2}

\usepackage{graphicx}% Include figure files
\usepackage{dcolumn}% Align table columns on decimal point
\usepackage{bm}% bold math
\usepackage{amssymb,amsfonts,amsmath}
\usepackage{listings,bbm,mathtools}
\usepackage{hyperref}% add hypertext capabilities
\hypersetup{colorlinks=true,linkcolor=blue,citecolor=blue}

\begin{document}

% \title{Supplemental Materials for \\ A Universal Matrix Ensemble that Unifies Eigenspectrum Laws \\via Neural Network Models}
% \title{A Universal Matrix Ensemble that Unifies Eigenspectrum Laws \\via Neural Network Models}
\title{Spectral density of correlated random matrices\\and nonmonotonic stability in hetero-associative memory networks}

\author{Arata Tomoto}
\email{tomoto.arata.37s@st.kyoto-u.ac.jp}
\author{Jun-nosuke Teramae}
\email{teramae@acs.i.kyoto-u.ac.jp}
\affiliation{Graduate School of Informatics, Kyoto University, Sakyo-ku, Kyoto 606-8502, Japan}
\date{\today}
\begin{abstract}
Random matrix theory, which characterizes spectral distributions of infinitely large matrices, plays a central role across diverse fields, including high-dimensional data analysis, ecology, neuroscience, and machine learning. Among its key results, the Marchenko–Pastur law and the elliptic law have provided theoretical foundations for numerous applications. However, despite their importance, the relationship between these two laws has not yet been fully understood. Here, we develop a novel derivation of the spectral density for a correlated random matrix ensemble that unifies the Marchenko–Pastur and elliptic laws as special cases. Interestingly, matrices from this ensemble can be naturally interpreted as connectivity matrices of hetero-associative memory networks, which, from a modern neural network perspective, are essentially equivalent to the linear attention architecture, a variant of the attention layer in the Transformer.
Using this result, we find that the stability of the network depends non-monotonically on the number of memorized patterns. By uncovering a nontrivial property of high-dimensional correlated systems, this work deepens our understanding of asymmetric interactions across various scientific fields.
% Random matrix theory, which characterizes the spectrum distribution of infinitely large matrices, plays a central role in theories across diverse fields, including high-dimensional data analysis, ecology, neuroscience, and machine learning. Among its celebrated achievements, the Marchenko--Pastur law and the elliptic law have served as key results for numerous applications. However, the relationship between these two laws remains elusive, and the existence of a universal framework unifying them is unclear. Inspired by a neural network model, we establish a universal matrix ensemble that unifies these laws as special cases. Through an analysis based on the saddle-node equation, we derive an explicit expression for the spectrum distribution of the ensemble. As a direct application, we reveal how the universal law clarifies the stability of a class of associative memory neural networks. By uncovering a fundamental law of random matrix theory, our results deepen the understanding of high-dimensional systems and advance the integration of theories across multiple disciplines.
\end{abstract}
\maketitle
\include{paper}
\clearpage
\include{supple}
\end{document}

%% file: paper.tex
Random matrix theory (RMT) investigates the universal properties of matrices with randomly generated elements \cite{brody1981random,mehta2004random,tao2012topics}. As a fundamental mathematical framework, it plays a crucial role in the theoretical analysis of diverse fields where large matrices naturally arise and serves as an essential tool for analyzing complex, high-dimensional systems. In particular, understanding the asymptotic behavior of eigenvalue distributions in the limit of infinite dimensions is a key issue in analyzing large matrices. Over the decades, extensive research has established fundamental laws governing these distributions, offering deep insights into universal principles that remain independent of the specific characteristics of individual matrices \cite{wigner1958distribution,marcenko1967distribution,girko1985circular,girko1986elliptic,sommers1988spectrum}. 

One of the most significant applications of RMT lies in high-dimensional data analysis \cite{wainwright2019high,pearson1901liii}. Consider an $N\times M$ data matrix $U$, where $N$ represents the number of $M$-dimensional data points. In the limit of large matrix sizes ($N, M \to\infty$), the eigenvalue distribution of the covariance matrix $C=UU^\top/M$ converges to the well-known "Marchenko–Pastur law," provided that the elements of the data matrix are independently and identically distributed (i.i.d.) \cite{marcenko1967distribution}. As a fundamental result for symmetric matrices, the Marchenko–Pastur law plays a central role in the analysis of extremely large datasets and provides the basis for various data analysis methods, including machine learning \cite{pennington2017nonlinear,adlam2022random} and dimensionality reduction techniques such as principal component analysis \cite{pearson1901liii}.

Another major application of RMT is in the study of dynamical systems, including neural networks and ecosystems \cite{amari1972characteristics, sompolinsky1988chaos, rajan2006eigenvalue,baron2020dispersal-induced,hu2022spectrum,baron2023breakdown}. A central issue in this field is the stability of a system under perturbations, which is determined by the largest eigenvalue and the spectral radius of the Jacobian matrix evaluated around a fixed point. In the limit of large system size, RMT predicts that the eigenvalue distribution of a generally asymmetric Jacobian matrix converges to the seminal "circular law," or more generally, to the "elliptic law" \cite{girko1985circular,girko1986elliptic,sommers1988spectrum}. These laws characterize the phase transition from a quiescent stable state to a dynamically chaotic regime, making them essential for analyzing large dynamical systems described by non-symmetric large matrices.

Despite their fundamental importance, relatively few studies have explored these two branches of limiting laws within a unified framework. Recently, however, seminal studies on the non-Hermitian generalization of the Marchenko–Pastur law have succeeded in deriving a remarkable result that a general eigenvalue distribution can recover the existing laws as special cases of it \cite{kanzieper2010non,akemann2021non,baron2025path}.

% In this study, we provide a novel derivation of the general random matrix ensemble that unifies these branches. In particular, we derive the eigenvalue density using an approach based on the eigenvalue potential, which is a representative approach for deriving the eigenvalue distribution of asymmetric random matrices \cite{sommers1988spectrum, haake1992statistics, rajan2006eigenvalue,kuczala2016eigenvalue, baron2020dispersal-induced, baron2022eigenvalue, baron2022eigenvalues,poley2024eigenvalue}.

In this study, we provide the general random matrix ensemble that unifies these branches for real matrices. In particular, we derive the eigenvalue density using an approach based on the eigenvalue potential, which is a representative approach for deriving the eigenvalue distribution of asymmetric random matrices \cite{sommers1988spectrum, haake1992statistics, rajan2006eigenvalue,kuczala2016eigenvalue, baron2020dispersal-induced, baron2022eigenvalue, baron2022eigenvalues,poley2024eigenvalue}. 
% Although it has the same density functional form, the result for real matrices does not follow as a trivial consequence of existing analyses for complex matrices.\cite{akemann2021non}.
In contrast to the previous derivation of the eigenspectrum, this approach allows us to derive the complex eigenspectrum of real non-symmetric Wishart matrices. Since real matrices are known to introduce non-trivial technical complications compared to complex matrices, the previous approach of \cite{akemann2021non} is not directly applicable to the real case. Although the real case was addressed only very recently \cite{byun2025real}, the analysis there is limited to real eigenvalues.

Importantly, the class of random matrices to which the law applies has a structure that can be interpreted as the synaptic connectivity matrix of one of the pioneering models of neural networks, that is, the hetero-associative memory model.
% This model stores a large set of input–output pattern pairs where each output is correlated with its corresponding input \cite{kohonen1972correlation,amari1972learning} and is also a generalization of the well-known Amari–Hopfield network \cite{amari1972learning,hopfield1982neural}.
This model stores a large set of input–output pattern pairs where each output is correlated with its corresponding input \cite{kohonen1972correlation,amari1972learning}. It is also a generalization of the well-known Amari–Hopfield network \cite{amari1972learning,hopfield1982neural}, and is essentially equivalent to a linear attention architecture, a variant of the attention layer that plays a central role in modern neural network architectures such as transformers.
To illustrate this point, we apply the derived eigenvalue distribution to identify the transition point of the network and reveal that it depends nontrivially on the number of stored memory patterns, demonstrating its relevance to real-world applications.

Consider an ensemble of random matrices $J = (J_{ij})$, defined as the product of two generally correlated random matrices:
\begin{align}
 J = \frac{1}{\sqrt{NM}}UV^\top.
\label{J_ij}
\end{align}
Here, $U=(U_{ij})$ and $V=(V_{ij})$ are $N\times M$ matrices whose components are Gaussian random variable with $\langle U_{ij}\rangle=\langle V_{ij}\rangle=0$, $\langle U_{ij} U_{kl}\rangle=\langle V_{ij} V_{kl}\rangle=\delta_{ik}\delta_{jl}$, and $\langle U_{ij} V_{kl}\rangle=\tau \delta_{ik}\delta_{jl}$, where $-1\leq \tau \leq 1$. Thus, the joint probability density function of $U$ and $V$ can be written as:
\begin{align}
    P(U, V) &= \frac{1}{(2\pi \sqrt{1-\tau^2})^{NM}}\notag\\
    \times &\exp\left[
                -\frac{1}{1-\tau^2}
                \operatorname{Tr}
                \left(
                    \frac{UU^\top+VV^\top}{2}-\tau UV^\top
                \right)
            \right]
    .
    \label{J_prob}  
\end{align}

Now, let us consider the eigenvalue distribution of the matrix $J$ in the complex plane for general values of $-1\leq\tau\leq 1$. Since $M$ determines the rank of $J$ and there exist $N-M$ trivial eigenvalues when $M<N$, the eigenspectrm density function is expressed as
\begin{align}
 \rho(\omega)
 &=
 \left\langle
  \frac{1}{N}\sum_{i=1}^{N} \delta(\omega-\lambda_i)
 \right\rangle_J \nonumber
 \\
 &=
 \rho_{\rm{b}}(\omega)
 + \left[1-\alpha\right]^{+} \delta(\omega)
\end{align}
where $\lambda_i$ denotes the $i$th eigenvalue of $J$, $\delta(x)$ is the Dirac delta function, and the $\langle\cdot\rangle_J$ indicate the ensemble average over the distribution given by Eq.~\eqref{J_prob}. Here, we define $\alpha=M/N$ and $[x]^{+}=\max(0, x)$. The term $\rho_{\rm{b}}(\omega)$ represents the contribution of the bulk regime to the density distribution, where the vast majority of eigenvalues are confined.

To derive the bulk density, following previous studies \cite{sommers1988spectrum, haake1992statistics, rajan2006eigenvalue,kuczala2016eigenvalue, baron2020dispersal-induced, baron2022eigenvalue, baron2022eigenvalues,poley2024eigenvalue}, we introduce the “potential” function $\Phi(\omega)$, defined over the complex plane except at points where $\omega$ coincides with one of the eigenvalues $\{\lambda_i\}$:
\begin{align}
 \Phi(\omega) &= - \frac{1}{N}
 \left<
  \log \det
  \left(
   (\omega^{*}\mathbbm{1}-J^\top)(\omega\mathbbm{1}-J)
  \right)
 \right>_{J},
\label{def_Phi}
\end{align}
where $\mathbbm{1}$ is the $N$-dimensional identity matrix, and the superscript $^{*}$ denotes complex conjugate. The derivative of the potential function is known to yield the Green's function, i.e., the disorder-averaged resolvent of $J$:
\begin{align}
 G(\omega)
 &= \frac{\partial\Phi}{\partial \omega}
 = \frac{1}{N}
 \left<
  \operatorname{Tr}\frac{1}{\omega\mathbbm{1}-J}
 \right>_J.
 \label{green}
\end{align}
Moreover, differentiating the Green's function with respect to $\omega^{*}$ provides the bulk density function as:
\begin{align}
 \rho_{\rm{b}}(\omega) = \frac{1}{\pi}\operatorname{Re}
 \left[
  \frac{\partial G}{\partial \omega ^*}
 \right].
 \label{rho_G}
\end{align}

% \begin{figure*}[htbp]
\begin{figure}[htbp]
  \includegraphics[width=\columnwidth]{"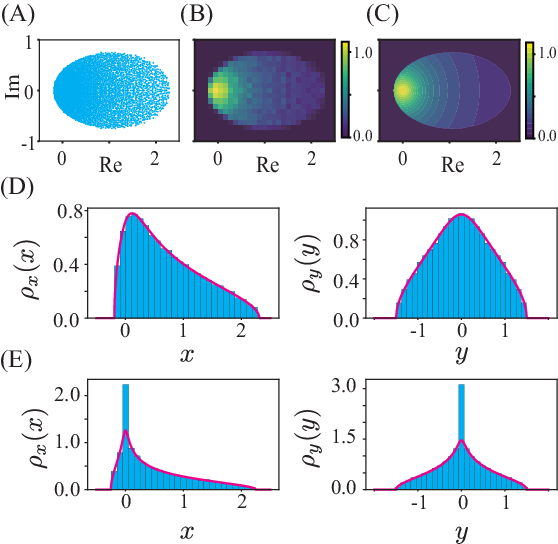"}
  \caption{Spectrum distribution of a product of large correlated random matrices. (A) Eigenvalues obtained numerically from a single realization of a randomly generated matrix from the ensemble defined by Eq.~\eqref{J_ij} and ~\eqref{J_prob}. Parameters are $\tau=0.5$, $N=5000$, and $M=10000$. Each dot represents an eigenvalue plotted in the complex plane. (B) Two-dimensional histogram of the eigenvalues shown in (A). (C) The theoretical prediction of the bulk density function, given by Eq.~\eqref{rho}, accurately describes the eigenvalue distribution. The distribution forms an elliptical region extracted from a function that is symmetric about the origin. (D) Marginalized distributions of eigenvalues onto the real axis (left panel) and the imaginary axis (right panel). Thick lines represent the theoretical predictions. (E) Same as (D), but for $M=4500$. Because $\alpha=M/N<1$, the matrix possesses trivial eigenvalues, which appear as a delta peak at the origin. Note that the theoretical prediction of the bulk density function precisely fits the distribution of the remaining eigenvalues.}
  \label{fig_check}
% \end{figure*}
\end{figure}

In the limit of large matrix sizes, $N, M\to\infty$, while keeping $M/N=\alpha$ fixed, the potential function $\Phi(\omega)$ can be evaluated by performing the ensemble average of the logarithm in Eq.~\eqref{def_Phi} using the saddle-point approximation \cite{supplementary}. This yields the Green's function as
\begin{align}
 G(\omega) = \frac{\tau\beta-E\omega^{*}}{\sigma E+1-\tau^2}
 .
 \label{G_equ}
\end{align}
Here, $\beta=\sqrt{\alpha}$, and $\sigma$ and $E$ are the solutions of the saddle-point equations:
\begin{align}
    \left\{
    \begin{aligned}
        \frac{\varepsilon}{\sigma^2}
        &=
        E - \frac{\left(Ex-\tau\beta\right)^2+\left(Ey\right)^2}{\left(1-\tau^2\right)^2}+O(\sigma) \\
        \frac{\varepsilon}{\sigma^2}
        &= \beta^2 E - \frac{\left(\tau Ex-\beta\right)^2+\left(\tau Ey\right)^2}{\left(1-\tau^2\right)^2}+O(\sigma)
    \end{aligned}
    \right.
    ,
    \label{varepsilon}
\end{align}
where $x + iy = \omega$, and $\varepsilon$ is an infinitesimal positive constant introduced to avoid singularities that arise when $\omega$ coincides with one of the eigenvalues $\lambda_i$.

In the limit of $\varepsilon\to 0^+$, the saddle-point equation admits two qualitatively distinct solutions \cite{supplementary}. One solution, characterized by $\sigma\to 0$, yields a nonzero bulk density within an elliptical region in the complex plane, while the other, with $\sigma>0$, ensures $\rho_{\rm{b}}=0$ outside this region. Consequently, we obtain the bulk density function as:
\begin{widetext}
\begin{equation}
  \rho_{\rm{b}}(\omega)=
  \begin{dcases}
    \frac{\beta}{2\pi(1-\tau^2)}\left(\left(\frac{1-\tau^2}{2}\left(\beta-\frac{1}{\beta}\right)\right)^2+x^2+y^2\right)^{-1/2}
    & \text{if } \left(\frac{x-\tau\left(\beta+\dfrac{1}{\beta}\right)}{1+\tau^2}\right)^2
    +\left(\frac{y}{1-\tau^2}\right)^2<1\\
    0 & \text{otherwise }
  \end{dcases},
  \label{rho}
\end{equation}
\end{widetext}
which reproduces the density function obtained in previous studies on the non-Hermitian generalization of the Marchenko–Pastur law\cite{kanzieper2010non,akemann2021non,baron2025path}.
A direct calculation confirms that the integral of the bulk density is correctly normalized \cite{supplementary} as
\begin{align}
  \int_{\mathbb{C}}\rho_{\rm{b}}(\omega)d\omega&=\frac{1+\alpha-\left|\alpha-1\right|}{2} \nonumber \\
  &=\min\left(1,\alpha\right)
  .
  \label{normalized}
\end{align}

% It is interesting to observe that the derived density function is symmetric about the origin, whereas its boundary contour is determined by an ellipse that is not necessarily centered at the origin. Also note that a direct calculation confirms that the integral of the bulk density is correctly normalized \cite{supplementary} as
% \begin{align}
%   \int_{\mathbb{C}}\rho_{\rm{b}}(\omega)d\omega&=\frac{1+\alpha-\left|\alpha-1\right|}{2} \nonumber \\
%   &=\min\left(1,\alpha\right)
%   .
%   \label{normalized}
% \end{align}
% % Although this result is obvious from the definition, it is still fascinating that an elliptical region of a function symmetric about the origin precisely satisfies this relationship.

\begin{figure}[htbp]
  \centering
  \includegraphics[width=8cm]{"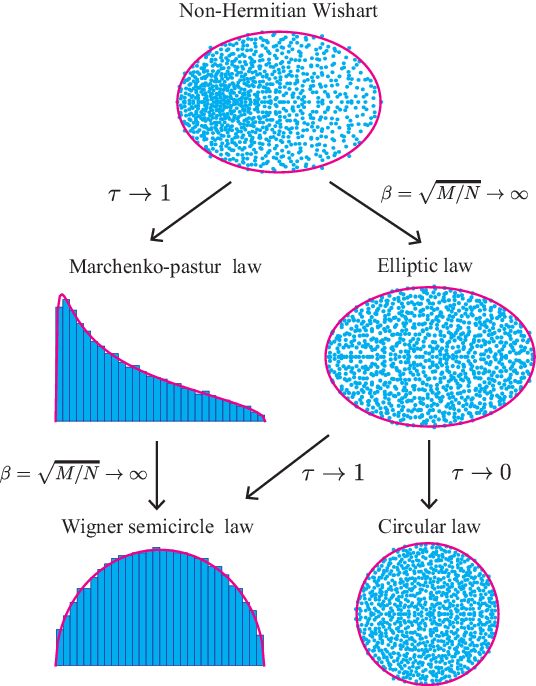"}
  \caption{The eigenspectrum density of products of large correlated random matrices, which is provided by \cite{kanzieper2010non,akemann2021non,baron2025path} (Eq.~\eqref{rho}), recovers the Marchenko–Pastur law and the elliptic law as special cases corresponding to specific limits of the ensemble’s parameters. Furthermore, from these two laws, other fundamental results of random matrix theory, such as the Wigner semicircle law and the circular law, can also be derived.}
  \label{fig_relation}
\end{figure}

To validate the theoretical results, we numerically evaluate the eigenvalues of a randomly generated matrix $J$ with $ N = 5000$. Figure \ref{fig_check} clearly shows that the theoretical prediction accurately explains the numerical results. As mentioned previously, we can also confirm both theoretically and numerically that the spectral density unifies two known eigenvalue distributions, namely the Marchenko–Pastur law and the elliptic law, by reproducing them as special cases \cite{supplementary} (Fig.~\ref{fig_relation}).

% To validate the theoretical results, we numerically evaluate the eigenvalues of a randomly generated matrix $J$ with $ N = 5000$. Figure \ref{fig_check} clearly shows that the theoretical prediction accurately explains the numerical results. Moreover, the results demonstrate the self-averaging property of the matrix ensemble, as the eigenspectrum of a single realization closely agrees with the theoretical expression for the spectrum, which is defined as the ensemble average. As mentioned previously, we can also confirm both theoretically and numerically that the spectral density unifies two known eigenvalue distributions, namely the Marchenko–Pastur law and the elliptic law, by reproducing them as special cases \cite{supplementary} (Fig.~\ref{fig_relation}).

Note that our results empirically exhibit the self-averaging property and universality, which are key properties observed in many random matrix ensembles, such as the Wigner–Dyson symmetry classes and Ginibre matrices \cite{wigner1958distribution,dyson1962statistical,mehta2004random,tao2010random,tao2011random,o2015products}. In the limit of large matrix sizes, the spectral distribution becomes independent of individual realizations and of the detailed form of the entry distributions, and depends only on a small number of parameters characterizing the low-order moments.

Interestingly, the matrix $J$, defined as the product of two correlated random matrices, can be regarded as the connectivity matrix of one of the most promising neural network models of associative memory, which includes the well-known Amari–Hopfield network \cite{amari1972learning,hopfield1982neural} as a special case. 
% This model has recently regained attention due to its structural similarity to the Transformer architectures\cite{}.
From a modern neural network perspective, this model can be considered essentially equivalent to a linear attention architecture, a variant of the attention layer in the Transformer.
In the rest of the paper, we explain this relationship. We show how the spectral distribution (Eq.~\eqref{rho}) reveals a nontrivial dependence of the stability transition point in a hetero-associative memory model on the number of stored memory patterns, demonstrating its practical relevance.

Memory retrieval in the brain associated with sensory inputs has been modeled as a network that stores multiple input-output pairs, namely key-value pairs, in its connectivity matrix \cite{kohonen1972correlation,amari1972learning}. Consider $M$ key-value pairs, where each key and value are represented as an $N$-dimensional random vector with zero mean and with correlation $\tau$ between each key and its corresponding value. Then, previous studies have shown that these key-value associations can be stored in the connectivity matrix $J$, defined as the superposition of the outer products of the key and value vectors \cite{kohonen1972correlation,amari1972learning}:
\begin{align}
J = \sum_{m=1}^{M} U^{(m)} V^{(m)\top},
\label{J_ij_memory}
\end{align}
where $U^{(m)}$ and $V^{(m)}$ denote the $m$th value and key vectors, respectively. When the system size $N$ is sufficiently large, and the number of embedded patterns $M$ satisfies suitable conditions, the network can stably recall the stored pattern $U^{(m)}$ corresponding to the input $V^{(m)}$ through the transformation $U^{(m)} = J V^{(m)}$. The model is referred to as the hetero-associative memory because $U^{(m)}$s and $V^{(m)}$s are generally different.

This model has a structure closely related to linear attention. For an input matrix Q, the network output is given by $UV^TQ = U(V^TQ)$. In this form, $Q$ and $V$ can be interpreted as analogs of the query and key matrices in linear attention after the application of feature maps, while $U$ plays the role of the value matrix\cite{katharopoulos2020transformers}.

To analyze this model, we introduce the correlation $\tau$ between $U^{(m)}$ and $V^{(m)}$. We then find that the connectivity matrix $J$ is identical to the matrix introduced in Eq.~\eqref{J_ij}, up to a scaling factor. This formulation also allows the connectivity matrix to reproduce the form of the well-known Amari-Hopfield auto-associative memory \cite{amari1972learning,hopfield1982neural}, in which each $U^{(m)}$ and $V^{(m)}$ are identical, in the limit $\tau \to 1$.

While the original hetero-associative model is defined by a one-step transformation from an input vector to its corresponding target vector, a slight modification of the connection matrix, still in the form of Eq.~\eqref{J_ij}, allows memory retrieval to be modeled as a recurrence relation. We can assume, without loss of generality, that the key-value pattern pairs are normalized such that $\|U^{(m)}\|=1$ and $\|V^{(m)}\|=k$, and that the correlation between them is $\tau$, that is, $\langle U^{(m)^\top} V^{(m)}\rangle = k\tau$. Then, denoting the state vector at time $t$ by $\bm{x}_t$, the network dynamics are given by
\begin{align}
\bm{x}_{t+1} &= r \bm{x}_t + \left(1-r\right) \left(J \bm{x}_t + V\right) \nonumber\\
J &\equiv \sum_{m=1}^{M} \left(U^{(m)}-V^{(m)}\right)U^{(m)\top},
\label{discrete}
\end{align}
where the scalar parameter $r$, satisfying $0<r<1$, represents the intrinsic decay rate of the neural state.

If the interference from other pattern pairs is negligible, the $m$th target vector $U^{(m)}$ becomes a fixed point of the dynamical system when the input $V$ is set to the corresponding key vector $V^{(m)}$. This formulation, in which the target pattern is represented as an input-dependent fixed point of the dynamics, is biologically more realistic and closer to the widely studied auto-associative memory model.

However, even though fixed points corresponding to the memorized key–value associations exist, the state vector $\bm{x}_t$ cannot converge to a stored pattern unless the corresponding fixed point is stable. Depending on the number of embedded pattern pairs, the network may lose stability, causing the trajectory to diverge rather than converge to a memorized pattern.

The stability of the dynamical system is given by the condition that the maximum magnitude of the eigenvalues of the matrix $r\mathbbm{1}+(1-r)J$ is smaller than unity. By applying the spectral density function in Eq.~\eqref{rho} to the connectivity matrix $J$ in Eq.~\eqref{discrete}, we obtain the stability condition of the fixed points of Eq.~\eqref{discrete} as
\begin{align}
\lambda &= \max\left(\left|r+\left(1-r\right)\lambda_{\pm}\right|\right) < 1.
\label{max_lambda}
\end{align}
Here,
\begin{align}
\lambda_{\pm} &= c\left(\beta^2+1\right)\pm\left(l+\frac{c^2}{l}\right)\beta
\end{align}
represent the left and right edges of the ellipse of the density function, where $l=\|U^{(m)} - V^{(m)}\|=\sqrt{k^2-2k\tau+1}$ and $c=1-k\tau$. As before, $\beta=\sqrt{M/N}$ is the square root of the ratio between the number of embedded pattern pairs $M$ and the number of neurons $N$. Note that only these edges on the real axis contribute to the maximum magnitude of the eigenvalues, since the real axis is always the major axis of the ellipse.

\begin{figure}[htbp]
  \includegraphics[width=\columnwidth]{"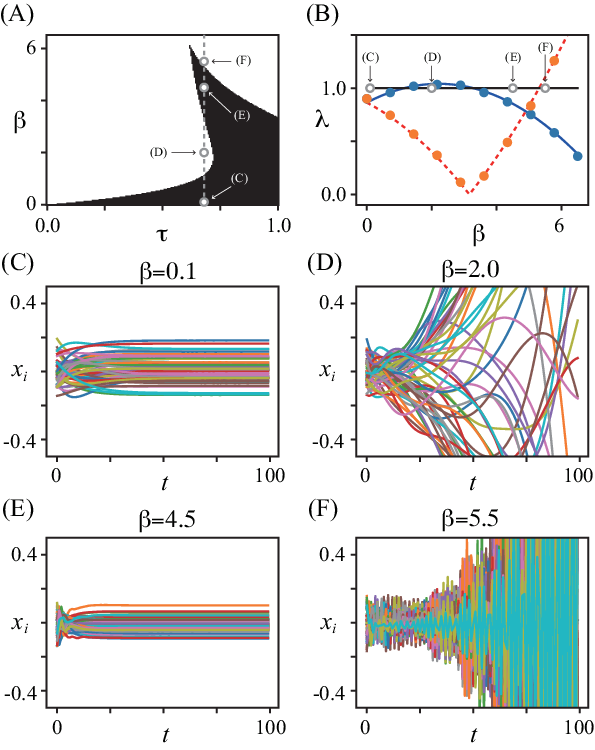"}
  \caption{Linear stability of the recurrent hetero-associative memory network whose connectivity matrix is defined by the sum of products of correlated key–value pairs (Eq.~\eqref{discrete}). (A) Phase diagram of the network stability in the $(\tau,\beta)$ plane. The shaded region indicates parameter values for which the network is linearly stable. The vertical gray dotted line indicates the value of $\tau$ used in (B)–(F). Gray circles in this panel and in panel (B) indicate values of $\beta$ used in (C)–(F). (B) Left and right candidates for the maximum eigenvalue, $\left|r+(1-r)\lambda_{+}\right|$ (thick blue line) and $\left|r+(1-r)\lambda_{-}\right|$ (red dashed line). Circles on the curves indicate results of numerical simulations using randomly generated matrices with $N=200$. The system is linearly stable when both curves lie below the horizontal black line at unity. (C–F) Examples of the temporal evolution of the network. As the number of embedded patterns increases, that is, as $\beta$ increases, the system undergoes repeated transitions between stable and unstable regimes. Randomly selected 50 components out of 200 are shown. The values of $\beta$ are 0.1 (C), 2.0 (D), 4.5 (E), and 5.5 (F). Other parameters are $k=2.0$ and $r=0.9$ in all panels. All numerical simulations were performed using the RK45 method with $N = 200$ \cite{hairer1993solving}.}
  % \caption{Linear stability of a recurrent neural network whose connectivity matrix is defined by the sum of correlated key-value pairs. (A) The largest eigenvalue of the Jacobian matrix. The dotted line shows the average over 100 realizations of the connectivity matrix; error bars indicate the standard deviation, and the thick line represents the theoretical prediction given by Eq.~\eqref{max_lambda}. (B) Numerically obtained phase diagram of the network’s stability for $\alpha = 1$. The shaded region indicates parameter values for which the network does not converge to the fixed point and is thus unstable. The thick line represents the theoretically derived phase boundary $\tau = 1/\sqrt{g} - 1$. (C) An example of the temporal evolution of the network in the stable regime ($g = 0.5$, $\tau = 0.25$). We randomly select 20 neurons out of 1000 neurons. (D) Same as (C), but in the unstable regime ($g = 0.7$, $\tau = 0.5$). All numerical simulations were performed using the RK45 method with $N = 1000$ \cite{hairer1993solving}.}
  \label{fig_dynamics}
\end{figure}

Interestingly, the stability of the dynamical system depends non-monotonically on the number of embedded pattern pairs and undergoes repeated transitions between stable and unstable regimes. Figure \ref{fig_dynamics} summarizes this dependence. As expected, the system loses its stability as the number of embedded pattern pairs increases. However, for certain parameter ranges with intermediate values of the correlation $\tau$, it reenters the stable region again before eventually losing stability and entering the unstable regime. This non-monotonicity arises from the competition between the left and right edges of the ellipse and their nonmonotonic dependence on $\beta$, which itself originates from the fact that the center of the ellipse depends on $\beta$ in a nontrivial manner as $\beta + 1/\beta$, as given in Eq.~\eqref{rho}.

In conclusion, we present a novel derivation based on the potential function of the spectral distribution for correlated large random matrices, which unifies the Marchenko–Pastur law and the elliptic law as special cases. To illustrate the significance of this ensemble and the applicability of the derived density function, we show that the resulting random matrix coincides with the connection matrix of a hetero-associative memory network that stores generally correlated pairs of key–value associations. By considering the recursive retrieval dynamics of the network, we further demonstrate that, counterintuitively, the stability of the fixed points depends non-monotonically on the number of embedded pattern pairs. This behavior reflects the dependence of the contour ellipse of the spectral density on the number of embedded patterns. Although we applied the framework only to a representative neural network for associative memory, correlated large systems with non-reciprocal connections are ubiquitous across multiple fields \cite{fruchart2021non,blumenthal2024phase,avni2025nonreciprocal}. This ubiquity suggests a natural next step of extending the framework to diverse dynamical systems, including complex ecological systems, cortical networks, and artificial neural networks such as the Transformer.
% In particular, linear attention mechanisms can be formulated as superpositions of key–value pairs, suggesting a relationship to the present framework and indicating that our results may be relevant to their stability and dynamics.
In particular, since linear attention mechanisms can be formulated as superpositions of many key–value pairs, the architecture of linear attention can be modeled by the connection matrix studied here. This suggests that our results may be relevant to the study of the stability and dynamics of such attention architectures in modern neural networks.
In general, systems with a directed input–output architecture are expected to exhibit nonreciprocal connectivity, with the hetero-associative memory network studied here acting as a representative example of such systems. Therefore, it is also a promising and important direction for future work to further analyze the properties of the nonmonotonic reentrant stability discovered in this study and to investigate its generality across a wide range of systems characterized by nonreciprocal connectivity arising from directed input–output architectures.

We thank Tomoki Fukai for valuable discussions and comments. J. T. is supported by JSPS KAKENHI Grant Number JP24K15104 and JP21H02597, and Japan Agency for Medical Research and Development (AMED) AMED-CREST JP25gm1510005s0205.

% In conclusion, we proved that the eigenvalue distribution of products of correlated large random matrices provides a universal framework that unifies the Marchenko--Pastur law and the elliptic law, enabling a systematic understanding of their relationship. These two laws emerge as special cases of the theory when the parameters of the spectrum distribution reach their limiting values. When applied to a neural network with hetero-associative memory, the framework reveals a nontrivial dependence of network stability on the number of memorized patterns. Although we applied the framework only to one of the simplest neural networks, the ubiquity of correlated large systems with non-reciprocal connections across multiple fields \cite{fruchart2021non,blumenthal2024phase,avni2025nonreciprocal} suggests that a natural next step is to extend it to diverse dynamical systems, including complex ecological systems, cortical networks, and artificial neural networks such as the Transformer. Extending the framework beyond the spectrum distribution, particularly toward the eigenvalue spacing distribution, offers a promising direction for future research.

\input{ref.bbl}

%% file: supple.tex
\onecolumngrid
\allowdisplaybreaks
\section{SUPPLEMENTAL MATERIALS}
\subsection{Analytical derivation of the bulk spectral density}
In this section, we explain the derivation of the bulk density function (Eq.~(9) in the main text) by evaluating the potential function (Eq.~(4) in the main text) using the saddle-point approximation and solving the corresponding saddle-point equation.

\subsubsection{Ensemble average}

Evaluating the potential function (Eq.~(4) in the main text) requires taking the ensemble average of a logarithmic function, which is generally difficult to compute. Fortunately, however, previous studies have shown that, in the context of the present problem, the averaging operation and the logarithmic function can be interchanged in the large system-size limit. (This can be confirmed using the replica method, which shows that replicas decouple in this limit: see previous studies [8, 15, 16, 38--42] for details.) By interchanging the averaging operation and the logarithm, representing the determinant as a Gaussian integral, and applying the Hubbard-Stratonovich transformation [47] to eliminate second-order terms of $J$ in the exponential function, we have
\begin{align}
  \exp(N\Phi(\omega)) &=
    \left<
        \int \prod_{ij}\frac{d^2z_idy_j^2}{\pi^2}\exp
        \left(
            -\sum_{i}(\varepsilon y_i^*y_i+z_i^*z_i)+i\sum_{ij}
            \left(
                y_i^*\left(\omega^*\delta_{ij}-J_{ji}\right)z_j+
                z_i^*\left(\omega\delta_{ij}-J_{ij}\right)y_j
            \right)
         \right)
    \right>_J
    \nonumber
\\
&=
  \int\prod_{ij}\frac{d^2z_idy_j^2}{\pi^2}\exp
    \left(
        \sum_{i}
        \left(
            -\varepsilon y_i^*y_i-z_i^*z_i+i\omega^*y_i^*z_i+i\omega y_iz_i^*
        \right)
    \right)\times 
    \left<
    \exp \left(-i\sum_{ij}
        \left(
            z_i^*y_j+z_iy_j^*
        \right)J_{ij}
    \right)
    \right>_J
    .
  \label{int_Phi}
\end{align}
Here, to avoid singularities that occur when $\omega$ coincides with one of the eigenvalues of $J$, we have added the first term in the integrand by introducing a positive infinitesimal constant $\varepsilon$. (Note that the derivation up to this point follows previous studies.) Although this constant is eventually eliminated by taking the limit $\varepsilon\to 0$, we will see that this infinitesimal plays an important role in the derivation of our main result. 

To perform the ensemble average of the above expression, we substitute the definition of $J$ (Eq.~(1) in the main text) and use the joint probability density function of $U$ and $V$ (Eq.~(2) in the main text). The average can then be rewritten as
\begin{align}
    &\left\langle
        \exp \left(
            -i\sum_{ij}\left(z_i^*y_j+z_iy_j^*\right)J_{ij}
        \right)
    \right\rangle_J \nonumber \\
    = &
    \left\langle
        \exp \left(
            -\frac{i}{\sqrt{NM}}\sum_{ij}\left(z_i^*y_j+z_iy_j^*\right)\sum_{p}U_{ip}V_{jp}
        \right)
    \right\rangle_{\{U,V\}}
    \nonumber \\
    = &
    \frac{1}{\left(2\pi\sqrt{1-\tau^2}\right)^{NM}}
    \int\prod_{i,j}
        \exp \left(
            -\frac{i}{\sqrt{NM}}\sum_{ij}\left(z_i^*y_j+z_iy_j^*\right)\sum_{p}U_{ip}V_{jp}
            \right)
    \nonumber \\ 
        &\hspace{12em}\times \exp \left(
            -\frac{1}{2\left(1-\tau^2\right)}\sum_{p}
            \left(
                U_{ip}^2+V_{ip}^2-2\tau U_{ip}V_{ip}
            \right)
        \right)
    \nonumber \\
    = &
    \frac{1}{\left(2\pi\sqrt{1-\tau^2}\right)^{NM}}
    \int\prod_{i,j}dU_idV_j
        \left(\exp \left(
            -\frac{i}{\sqrt{NM}}\sum_{ij}\left(z_i^*y_j+z_iy_j^*\right)U_{i}V_{j}
            \right)\right.
    \nonumber \\ 
        &\hspace{12em}\times \exp \left.\left(
            -\frac{1}{2\left(1-\tau^2\right)}
            \left(
                U_{i}^2+V_{i}^2-2\tau U_{i}V_{i}
            \right)
        \right)
        \right)^M
    \nonumber\\
    = &
    \frac{1}{\left(2\pi\sqrt{1-\tau^2}\right)^{NM}}
    \left(
        \int \prod_{i,j}dU_idV_j
        \exp \left(
            -\frac{1}{2}
            \left(U\ \  V\right)
            \begin{pmatrix}
                Q_{UU} & Q_{UV} \\
                Q_{VU} & Q_{VV}
            \end{pmatrix}
            \begin{pmatrix}
                U \\ V
            \end{pmatrix}
        \right)
    \right)^M \nonumber \\
    = &
    \frac{1}{\left(\left(1-\tau^2\right)^N\det Q\right)^{M/2}}
    .
    \label{gaussian_average}
\end{align}
To obtain the fourth line, we used the fact that $U_{ip}$ and $V_{ip}$ for different $p$ are mutually independent, and omitted the subscript $p$ in the line, because all terms corresponding to different $p$ contribute equally to the exponential function. We introduced the following notation in the final two lines:
\begin{align}
    & Q \equiv 
    \begin{pmatrix}
        Q_{UU} & Q_{UV} \\
        Q_{VU} & Q_{VV}
    \end{pmatrix} \\
    & Q_{UU} = Q_{VV} = \frac{1}{1-\tau^2}\mathbbm{1} \\
    & Q_{UV} =
        -\frac{\tau}{1-\tau^2}\mathbbm{1}
        + \frac{i}{\sqrt{NM}}\left(y\left(z^*\right)^\top+y^*z^\top\right) \\
    & Q_{VU} = 
        -\frac{\tau}{1-\tau^2}\mathbbm{1}
        +\frac{i}{\sqrt{NM}}\left(z\left(y^*\right)^\top+z^*y^\top\right)
        ,
\end{align}
where $y=(y_1, \ldots, y_N)^{\top}$ and $z=(z_1, \ldots, z_N)^{\top}$ are $N$-dimensional vectors. In the last line, we performed the Gaussian integrals over $\{U_i\}$ and $\{V_i\}$.

The determinant $\det Q$ is evaluated as follows. First, by applying the block matrix formula, we obtain  
\begin{align}
    \det Q
    &= \det Q_{UU} \det (Q_{VV}-Q_{VU}Q_{UU}^{-1}Q_{UV}) \nonumber \\  
    &= \frac{1}{\left(1-\tau^2\right)^N}\det \tilde{Q},
\end{align}
where we defined $\tilde{Q}\equiv Q_{VV}-Q_{VU}Q_{UU}^{-1}Q_{UV}$. Next, we observe that terms proportional to $yy$, $zz$, $yz$, and their complex conjugates are absent in Eq.\eqref{int_Phi}, which implies that they do not contribute to $\det \tilde{Q}$[38]. Thus, by direct calculation with excluding these terms, we obtain:
\begin{align}  
    \tilde{Q}
    \begin{pmatrix}
      y \\ z
    \end{pmatrix}
    =\begin{pmatrix}
      1+\dfrac{i\tau z^*y}{\sqrt{NM}} & \dfrac{i\tau y^*y}{\sqrt{NM}}+\dfrac{(1-\tau^2)y^*y\cdot z^*y }{NM} \\
      \dfrac{i\tau z^*z}{\sqrt{NM}} & 1+\dfrac{i\tau y^*z}{\sqrt{NM}}+\dfrac{(1-\tau^2)y^*y\cdot z^*z}{NM} 
    \end{pmatrix}
    \begin{pmatrix}
      y \\ z
    \end{pmatrix}
    \equiv\tilde{Q}_{yz}
    \begin{pmatrix}
      y \\ z
    \end{pmatrix}
    .
    \label{Q_yz}
\end{align}  
This indicates that $y$ and $z$ span a linear subspace on which $\tilde{Q}$ acts as $\tilde{Q}_{yz}$, the restriction of $\tilde{Q}$ to that subspace. Thus, this subspace contributes a factor $\det \tilde{Q}_{yz}$ to $\det \tilde{Q}$. Similarly, one can show that $y^*$ and $z^*$ span another subspace on which $\tilde{Q}$ acts in the same way, and this subspace yields the same contribution to the determinant. In addition, for any vector $x$ that is orthogonal to both of these subspaces, we have $\tilde{Q}x = x$. This implies that the remaining $(N-4)$-dimensional orthogonal subspace contributes a factor of $1$ to $\det \tilde{Q}$.

Combining these contributions, we obtain  
\begin{align}  
    \det \tilde{Q}
    &=
    \left(\det \tilde{Q}_{yz}\right)^2
    =
    \left(
        \frac{y^*y\cdot z^*z}{NM} +
        \left(
            1+\frac{i\tau y^*z}{\sqrt{NM}}
        \right)
        \left(
            1+\frac{i\tau z^*y}{\sqrt{NM}}
        \right)
    \right)^2
    ,
  \label{det_tilde_Q}
\end{align}  
which leads to  
\begin{align}
    \left\langle
        \exp \left(
            -i\sum_{ij}\left(z_i^*y_j+z_iy_j^*\right)J_{ij}
        \right)
    \right\rangle_J
    &= 
    \left(
        \frac{y^*y\cdot z^*z}{NM} +
        \left(
            1+\frac{i\tau y^*z}{\sqrt{NM}}
        \right)
        \left(
            1+\frac{i\tau z^*y}{\sqrt{NM}}
        \right)
    \right)^{-M}
    .
    \label{average}
\end{align}

\subsubsection{Saddle-point approximation}

Substituting the average Eq.~\eqref{average} to Eq.~\eqref{int_Phi} gives:
\begin{align}
    \exp\left(N \Phi\left(\omega\right)\right) &=
    \int\prod_{ij}\frac{d^2z_idy_j^2}{\pi^2}
    \exp
    \left[N\left\{
        -\varepsilon \frac{y^*y}{N}-\frac{z^*z}{N}+i\omega^*\frac{y^*z}{N}+i\omega \frac{z^*y}{N} 
    \right\}\right] \nonumber \\
        &\hspace{7em}\times \exp\left[-M\log
        \left(
            \frac{y^*y\cdot z^*z}{NM} +
            \left(
                1+\frac{i\tau y^*z}{\sqrt{NM}}
            \right)
            \left(
                1+\frac{i\tau z^*y}{\sqrt{NM}}
            \right)
        \right)
    \right]
    \nonumber \\
    &=
    \int\prod_{ij}\frac{d^2z_idy_j^2}{\pi^2}
    \exp\left[N\left\{
        -\varepsilon \frac{y^*y}{N}-\frac{z^*z}{N}+i\omega^*\frac{y^*z}{N}+i\omega \frac{z^*y}{N} \right.\right.
        \nonumber \\
        &\hspace{12em}\left.\left.-\alpha \log
        \left(
            \frac{y^*y\cdot z^*z}{NM} +
            \left(
                1+\frac{i\tau y^*z}{\sqrt{NM}}
            \right)
            \left(
                1+\frac{i\tau z^*y}{\sqrt{NM}}
            \right)
        \right)
    \right\}\right]
    .
\end{align}
Since the exponent of the exponential function on the right-hand side is proportional to the system size $N$, we can evaluate the integral in the limit $N\to \infty$ using the saddle-point approximation.

Observe that the exponent is a function of only $y^* y$, $y^* z$, $z^* y$, and $z^* z$, so we introduce macroscopic order parameters defined by $A \equiv\sum_{i}y_i^*y_i/N$, $B \equiv \sum_{i}y_i^*z_i/N$, $C \equiv \sum_{i}z_i^*y_i/N$, and $D \equiv \sum_{i}z_i^*z_i/N$, and impose these definitions by inserting the Dirac delta functions into the integral:
\begin{align}
    \int dX\exp&
    \left(N
        \left(
            -\varepsilon A-D+i\omega^* B+i\omega C+\alpha\log
            \left(
                \frac{1}{\alpha}AD+
                \left(1+\frac{i\tau B}{\beta}\right)
                \left(1+\frac{i\tau C}{\beta}\right)
            \right)
        \right)
    \right) \nonumber \\
    &\times \int\prod_{ij}\frac{d^2z_idy_j^2}{\pi^2}
    \delta \left(y^*y-NA\right)
    \delta \left(y^*z-NB\right)
    \delta \left(z^*y-NC\right)
    \delta\left(z^*z-ND\right).
\end{align}
Here, for brevity, we write $X\equiv \{A, B, C, D\}$ and $dX=dAdBdCdD$. Then, we use the Fourier representations of the delta functions. For instance,
\begin{align}
    \delta\left(y^*y-NA\right)=
    \frac{1}{\left(2\pi\right)^N}\int d\hat{A} \exp
    \left(-i\hat{A}\left(y^*y-NA\right)\right),
\end{align}
and after performing the Gaussian integrals over $y$ and $z$, we obtain:
\begin{align}
    \exp\left(N \Phi\left(\omega\right)\right)&=
    \int dXd\hat{X} \exp
    \left(N
        \left(
            -\varepsilon A-D+i\omega^* B+i\omega C+\alpha\log
            \left(
                \frac{1}{\alpha}AD+
                \left(1+\frac{i\tau B}{\beta}\right)
                \left(1+\frac{i\tau C}{\beta}\right)
            \right)
        \right)
    \right) \nonumber \\
    &\hspace{5em} \times \frac{1}{\left(2\pi\right)^{4N}}\exp\left(iN\left(A\hat{A}+B\hat{B}+C\hat{C}+D\hat{D}\right)\right)
    \int\prod_{ij}\frac{d^2z_idy_j^2}{\pi^2} 
    \exp \left(-
        \left(y\ \  z\right)
        \begin{pmatrix}
            i\hat{A} & i\hat{B} \\
            i\hat{C} & i\hat{D}
        \end{pmatrix}
        \begin{pmatrix}
            y^* \\ z^*
        \end{pmatrix} 
    \right)\nonumber \\
    &= \int dXd\hat{X} \exp
    \left(N
        \left(
            -\varepsilon A-D+\omega^* B+\omega C+\alpha\log
            \left(
                \frac{1}{\alpha}AD+
                \left(1+\frac{\tau B}{\beta}\right)
                \left(1+\frac{\tau C}{\beta}\right)
            \right)
        \right)
    \right) \nonumber \\
    &\hspace{5em}\times 
    \exp\left(
        N\left(
            A\hat{A}+B\hat{B}+C\hat{C}+D\hat{D}
            -\log\left(\hat{A}\hat{D}+\hat{B}\hat{C}\right)
            -4\log\left(2\pi\right)
        \right)
    \right) \nonumber \\
    &\equiv \int dXd\hat{X} \exp
    \left(N
        \left(
            T(X)+S(X,\hat{X})
        \right)
    \right)
    \label{F_X}
    .
\end{align}
where $\hat{X}\equiv \{\hat{A}, \hat{B}, \hat{C}, \hat{D}\}$, and in the third line, we have replaced $iB,\ iC,\ i\hat{A},\ i\hat{D}$ with $B,\ C,\ \hat{A},\ \hat{D}$ to simplify the notation. In the last line, we defined two functions: $T$ as a function of $X$, and $S$ as a function of both $X$ and $\hat{X}$.

Applying the saddle-point approximation to Eq.~\eqref{F_X} yields the potential function as
\begin{align}
\Phi(\omega) = \operatorname{extr}_{X,\hat{X}}\left(T(X)+S(X,\hat{X})\right)
,
\label{saddle_node} 
\end{align}
where the right-hand side denotes the value of $T(X)+S(X,\hat{X})$ evaluated at its global maximum. Therefore, the saddle-point equation, that is, the stationary conditions for the eight variables $X = \{A, B, C, D\}$ and $\hat{X}=\{\hat{A}, \hat{B}, \hat{C}, \hat{D}\}$, are given by:
\begin{equation}
  \begin{aligned}
    &\frac{\partial F}{\partial A}=-\varepsilon -\frac{D}{E}+\hat{A}=0
    &&\frac{\partial F}{\partial \hat{A}}=  A-\frac{\hat{D}}{\hat{A}\hat{D}+\hat{B}\hat{C}}=0\\
    &\frac{\partial F}{\partial B}= \omega^*-\frac{\tau\left(\beta+\tau C\right)}{E}+\hat{B}=0
    &\qquad&\frac{\partial F}{\partial \hat{B}}=  B-\frac{\hat{C}}{\hat{A}\hat{D}+\hat{B}\hat{C}}=0\\
    &\frac{\partial F}{\partial C}=  \omega-\frac{\tau\left(\beta+\tau B\right)}{E}+\hat{C}=0
    &&\frac{\partial F}{\partial \hat{C}}=  C-\frac{\hat{B}}{\hat{A}\hat{D}+\hat{B}\hat{C}}=0\\
    &\frac{\partial F}{\partial D}= -1+\frac{A}{E}+\hat{D}=0
    &&\frac{\partial F}{\partial \hat{D}}=  D-\frac{\hat{A}}{\hat{A}\hat{D}+\hat{B}\hat{C}}=0
  \end{aligned}
  \label{saddle_equation}
\end{equation}
Here, for simplicity, we define a variable $E$ as:
\begin{align}
    E = \frac{1}{\alpha}AD
    + \left(1+\frac{\tau B}{\beta}\right)
    \left(1+\frac{\tau C}{\beta}\right).
\end{align}

\subsubsection{Solution of the saddle-point equations}

The right equations of the saddle-point equations Eq.~\eqref{saddle_equation} give the relation $\frac{\hat{D}}{A}=\frac{\hat{C}}{B}=\frac{\hat{B}}{C}=\frac{\hat{A}}{D}$, which allows us to eliminate the hat variables $\hat{X}$ from the equations. Then, by eliminating $B$, $C$, and $D$ using the left equations, we reduce the saddle-point equations to a two-dimensional equation:
\begin{equation}
  \left\{
    \begin{aligned}
      E&=\frac{\varepsilon}{\alpha\sigma^2}+ \left(1+\frac{\tau}{\beta}\cdot\frac{\tau\beta-E\omega^*}{\sigma E+1-\tau^2}\right)
      \left(1+\frac{\tau}{\beta}\cdot\frac{\tau\beta-E\omega}{\sigma E+1-\tau^2}\right) \\
      \sigma&=\varepsilon\left(1+\frac{1}{\sigma E}\right)+\sigma\left(\sigma+\frac{1}{E}\right)\cdot\frac{\left(Ex-\tau\beta\right)^2+(Ey)^2}{\left(\sigma E+1-\tau^2\right)^2}
    \end{aligned}
  \right.
,
  \label{saddle_point1}
\end{equation}
where $\sigma=\frac{1}{A}$. The remaining variables other than $A$ can then be expressed in terms of $\sigma$ and $E$ as:
\begin{equation}
  \begin{aligned}
    &B=\frac{\tau\beta-E\omega}{\sigma E+1-\tau^2},\quad C=\frac{\tau\beta-E\omega^*}{\sigma E+1-\tau^2},\quad D=\frac{\varepsilon}{\sigma} \\
    &\hat{A}=\varepsilon\left(1+\frac{1}{\sigma E}\right),\quad \hat{B}=\left(\sigma+\frac{1}{E}\right)C,\quad \hat{C}=\left(\sigma+\frac{1}{E}\right)B,\quad \hat{D}=1+\frac{1}{\sigma E}.
  \end{aligned}
  \label{saddle_solution}
\end{equation}

By solving each component of the two-dimensional saddle-point equation with respect to $\varepsilon$ and expanding them as a power series of $\sigma$, we obtain:
\begin{equation}
  \left\{
  \begin{aligned}
    \frac{\varepsilon}{\sigma^2} &= f(E) + O(\sigma) \\
    \frac{\varepsilon}{\sigma^2} &= g(E) + O(\sigma)
  \end{aligned}
  \right.
,
  \label{saddle_point2}
\end{equation}
where $O(\sigma)$ denotes terms of order $\sigma$ or higher, and the functions $f(E)$ and $g(E)$, which are independent of $\sigma$, are defined as:
\begin{align}
    f(E) &= E-\frac{\left(Ex-\tau\beta\right)^2+\left(Ey\right)^2}{\left(1-\tau^2\right)^2}
    \\
    g(E) &= \beta^2 E - \frac{\left(\tau Ex-\beta\right)^2+\left(\tau Ey\right)^2}{\left(1-\tau^2\right)^2}
    .
\end{align}
This form of the saddle-point equation implies that, in the limit of $\varepsilon \to 0^+$, the solution can take one of two forms. In the first case, $\varepsilon/\sigma^2$ remains finite as $\varepsilon \to 0^+$. In the second case, $\varepsilon/\sigma^2 \to 0$ in that limit.

The first case implies that $\sigma$ converges to $0$, with satisfying $\sigma \approx \sqrt{\varepsilon}$. Therefore, considering that $\sigma^2>0$, the saddle-point equation in this case reduces to a combination of an equality and an inequality:
\begin{align}
f(E) = g(E) > 0.
\end{align}
Solving the equation yields
\begin{align}
    E
    &= \frac{1}{2(x^2+y^2)}
    \left(
        \left(1-\beta^2\right)\left(1-\tau^2\right)
        + 2\beta\left(
            \left(\frac{1-\tau^2}{2}\left(\beta-\frac{1}{\beta}\right)\right)^2
            + x^2 +y^2
        \right)^{-1/2}
    \right) \nonumber \\
    &= \frac{1}{2\omega^*\omega}
    \left(
        \left(1-\beta^2\right)\left(1-\tau^2\right)
        + 2\beta\left(
            \left(\frac{1-\tau^2}{2}\left(\beta-\frac{1}{\beta}\right)\right)^2
            + \omega^*\omega
        \right)^{-1/2}
    \right)
    ,
\label{E}
\end{align}
and substituting this solution into the definitions of $f$ and $g$, we obtain the condition for the first case as the solution of the inequality:
\begin{align}
    \left( \frac{x-\tau\left(\beta+\frac{1}{\beta}\right)}{1+\tau^2}\right)^2
    +\left(\frac{y}{1-\tau^2}\right)^2<1 \label{elipse}
.
\end{align}
This condition defines the interior of an ellipse in the complex plane of $\omega=x+iy$.

When $\omega=x+iy$ satisfies the condition, the remaining variables other than $E$ of the solution can be obtained by substituting Eq.~\eqref{E} into Eq.~\eqref{saddle_solution}. Then, noting that $G=C$, which follows from Eq.~(6) in the main text and Eq.~\eqref{saddle_node}, we arrive at the bulk density function:
\begin{align}
     \rho_{\rm{b}}(\omega)& = \frac{1}{\pi}\operatorname{Re}
         \left[
          \frac{\partial C}{\partial \omega ^*}
         \right] 
     \nonumber \\
     &= \frac{1}{\pi}\operatorname{Re}
        \left[
            \frac{\partial}{\partial \omega ^*}
            \left(
            \frac{\tau\beta-E\omega^*}{\sigma E+1-\tau^2}
            \right)
        \right]
     \nonumber \\
     &=\frac{1}{\pi}\operatorname{Re}
        \left[
            \frac{\partial}{\partial \omega ^*}
            \left(
            -\frac{1}{2\omega(1-\tau^2)}
            \left(
              \left(
                1-\beta^2\right)\left(1-\tau^2\right)+2\beta\left(\left(\frac{1-\tau^2}{2}\left(\beta-\frac{1}{\beta}\right)\right)^2+\omega^*\omega\right)^{-1/2}
              \right)
              \right)
        \right]
     \nonumber \\
     &=
     \frac{\beta}{2\pi(1-\tau^2)}\left(\left(\frac{1-\tau^2}{2}\left(\beta-\frac{1}{\beta}\right)\right)^2+\omega^*\omega\right)^{-1/2}
     \nonumber \\
     &=\frac{\beta}{2\pi(1-\tau^2)}\left(\left(\frac{1-\tau^2}{2}\left(\beta-\frac{1}{\beta}\right)\right)^2+x^2+y^2\right)^{-1/2}
.
\end{align}
Note that this density function is symmetric with respect to the origin, whereas the boundary contour of the density function is given by the ellipse Eq.~\eqref{elipse}.

In the second case, where $\omega$ is outside the ellipse, $\sigma$ remains finite. Then, by substituting $\varepsilon=0$ into Eq.~\eqref{saddle_point1}, the saddle-point equation reduces to
\begin{align}
  \left\{
  \begin{aligned}
      E&=\left|
      1+\frac{\tau}{\beta}C
      \right|^2 \\
      1&=\frac{(\tau \beta-E\omega)C+\tau^2\left|C\right|^2}{E}.
  \end{aligned}
  \right.
\end{align}
Solving this equation yields the Green’s function as:
\begin{align}
    G(\omega)
    = C
    = \frac{\tau^2\beta-\beta\omega-\tau\pm\sqrt{(\tau^2\beta-\beta\omega-\tau)^2-4\tau\beta\omega}}{2\tau\omega}
.
\end{align}

Since the Green’s function depends only on $\omega$ and not on $\omega^*$, we conclude from Eq.~(6) in the main text that the bulk density vanishes outside the ellipse:
\begin{align}
\rho_{b}\left(\omega\right)&=0
.
\end{align}

Finally, by combining the above results, we obtain the full expression of the bulk density function:
\begin{align}
  \rho_{\rm{b}}(\omega)=
  \begin{dcases}
    \frac{\beta}{2\pi(1-\tau^2)}\left(\left(\frac{1-\tau^2}{2}\left(\beta-\frac{1}{\beta}\right)\right)^2+x^2+y^2\right)^{-1/2}
    & \text{if } \left(\frac{x-\tau\left(\beta+\dfrac{1}{\beta}\right)}{1+\tau^2}\right)^2
    +\left(\frac{y}{1-\tau^2}\right)^2<1\\
    0 & \text{otherwise }
  \end{dcases},
  \label{rho_supple}
\end{align}
which is Eq.~(9) in the main text.

\subsection{Confirmation of the normalization of the bulk density function}

In this section, we confirm that the derived bulk density function indeed satisfies the normalization condition:
\begin{align}
  \int_{\mathbb{C}}\rho_{\rm{b}}(\omega)d\omega 
&=
\int_{-\infty}^{\infty} \int_{-\infty}^{\infty} \rho_{\rm{b}}(x+iy)dxdy \nonumber
\\
&= \frac{1+\alpha-\left|\alpha-1\right|}{2} \nonumber \\
  &=\min\left(1,\alpha\right)
.
\end{align}
As we mentioned in the main text, when $M<N$, i.e., $\alpha=M/N<1$, the matrix $J$ has $N-M$ trivial eigenvalues. Therefore, the bulk density function must account for the distribution of the remaining $M$ nontrivial eigenvalues, whose total density is given by $\alpha$.

First, we marginalize the density function with respect to $y$ for fixed $x$ to obtain
\begin{align}
\int \rho_{\rm{b}}(\omega) dy
&=\int_{-Y}^{Y} \rho_{\rm{b}}(x+iy) dy \nonumber
\\
&=
\frac{\beta\left(1+\tau^2\right)}{2\pi\left(1-\tau^2\right)}\left[\tanh^{-1}\left(\frac{y}{\sqrt{k^2+x^2+y^2}}\right)\right]_{-Y}^{Y} \nonumber \\
&=\frac{\beta\left(1+\tau^2\right)}{2\pi\left(1-\tau^2\right)}\left(\log\left(1+\frac{Y}{\sqrt{k^2+x^2+Y^2}}\right) 
-\log\left(1-\frac{Y}{\sqrt{k^2+x^2+Y^2}}\right)\right) \label{y}
,
\end{align}
where, for brevity,  we have introduced $k^2\equiv\left(\left(\frac{1-\tau^2}{2}\right)\left(\beta-\frac{1}{\beta}\right)\right)^2$, $Y\equiv(1-\tau^2)\sqrt{1-\left(\frac{x-\tau\left(\beta+\frac{1}{\beta}\right)}{1+\tau^2}\right)^2}$.

Then, integrating the marginalized expression with respect to $x$ yields
\begin{align}
\int_{\mathbb{C}}\rho_{\rm{b}}(\omega)d\omega &=
\int_{\mu-(1+\tau^2)}^{\mu+(1+\tau^2)}
\int_{-Y}^{Y} \rho_{\rm{b}}(x+iy)dy dx \nonumber
\\
&=\frac{\beta\left(1+\tau^2\right)}{2\pi\left(1-\tau^2\right)}
\int_{-1}^{1}\left(\log\left(1+\frac{(1-\tau^2)\sqrt{1-x^2}}{2\tau x+k}\right)
  -\log\left(1-\frac{(1-\tau^2)\sqrt{1-x^2}}{2\tau x+k}\right)\right)dx \nonumber \\
&=\frac{\beta\left(1+\tau^2\right)}{2\pi\left(1-\tau^2\right)}
\int_{-1}^{1} \left(\log\left(x+a\sqrt{1-x^2}+b\right)-\log\left(x-a\sqrt{1-x^2}+b\right)\right)dx  \nonumber \\
  &=\frac{\beta\left(1+\tau^2\right)}{2\pi\left(1-\tau^2\right)}
  \left(
      \int_{-\frac{\pi}{2}-t_0}^{\frac{\pi}{2}-t_0}\frac{\sin^2{t}\cos{t_0}}{c+\cos{t}}dt+
    \int_{-\frac{\pi}{2}+t_0}^{\frac{\pi}{2}+t_0}\frac{\sin^2{t}\cos{t_0}}{c-\cos{t}}dt
  \right)
 \nonumber \\
  &=\frac{4\beta}{\pi}
  \left(
  \int_{T_0}^{-\frac{1}{T_0}} \frac{u^2}{\left((c-1)u^2+c+1\right)\left(u^2+1\right)^2}du
  +\int_{\frac{1}{T_0}}^{-T_0} \frac{u^2}{\left((c+1)u^2+c-1\right)\left(u^2+1\right)^2}du
  \right)
.
\end{align}
Here, we performed a change of variables $x\to (x-\mu)/(1-\tau^2)$ in the second line, defined $a=(1-\tau^2)/(2\tau)$, $b=k/(2\tau)$, and $c=b/{\sqrt{a^2+1}}$ in the third line, and introduced an angular variable $t_0$ defined by $\cos{t_0}=a/\sqrt{a^2+1}$ with the substitution $x=\sin(t\pm t_0)$ in the fourth line. In the last line, we also applied an additional variable transformation by introducing $u=\tan(t/2)$ and $T_0=\frac{\tan(t_0/2)+1}{\tan(t_0/2)-1}$. By applying the partial fraction decomposition and using the identity $\tan^{-1}{x}+\tan^{-1}{\frac{1}{x}}=\operatorname{sign}(x)\frac{\pi}{2}$, we arrive at
\begin{align}
  \int_{\mathbb{C}}\rho_{\rm{b}}(\omega)d\omega
  &=\frac{4\beta}{\pi}\left(
        \int_{T_0}^{-\frac{1}{T_0}}
        \left(
            \frac{c+1}{4(u^2+1)}-\frac{c^2-1}{4\left((c-1)u^2+c+1\right)}-\frac{1}{2(u^2+1)}
        \right)du\right.\nonumber \\
        &\hspace{4em}+\left.
        \int_{\frac{1}{T_0}}^{-T_0}
        \left(
            \frac{c-1}{4(u^2+1)}-\frac{c^2-1}{4\left((c+1)u^2+c-1\right)}+\frac{1}{2(u^2+1)}
        \right)du
    \right)\nonumber \\
  &=\frac{4\beta}{\pi}
      \left[
        \frac{c}{2}\tan^{-1}(u)-\frac{\sqrt{c^2-1}}{4}\left(\tan^{-1}\left(\sqrt{\frac{c-1}{c+1}}u\right)+\tan^{-1}\left(\sqrt{\frac{c+1}{c-1}}u\right)\right)
    \right]_{T_0}^{-\frac{1}{T_0}} \nonumber \\
  &= \frac{4\beta}{\pi}\cdot \frac{\pi}{4}\left(c-\sqrt{c^2-1}\right) \nonumber \\
  &=\frac{\beta^2+1}{2}-\sqrt{\frac{\left(\beta^2+1\right)^2-4\beta^2}{4}} \nonumber \\
  &=\frac{\alpha+1-\left|\alpha-1\right|}{2} \nonumber \\
  &=\min \left(1,\alpha\right)
,
\end{align}
which is the desired result.

\subsection{Convergence of the derived spectrum density to the Marchenko–Pastur law and the elliptic law}
In this section, we confirm that the Marchenko–Pastur law and the elliptic law are recovered as special cases of the spectrum density function (Eq.~\eqref{rho_supple}, or Eq.~(9) in the main text).

\begin{figure}[t]
  \includegraphics[width=\columnwidth]{"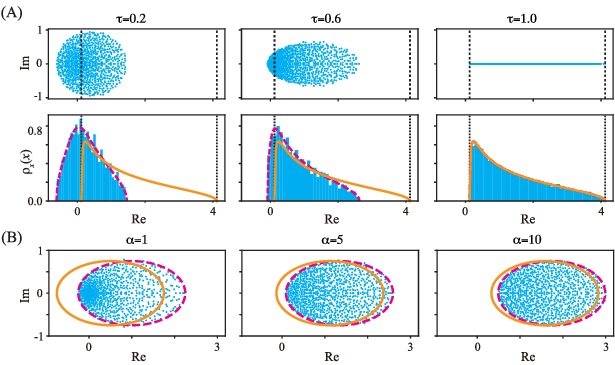"}
  \caption{Reduction of the obtained law to the marchenko–pastur and elliptic laws in limiting cases of the ensemble parameters. (A) As the parameter $\tau$ approaches unity, $J$ becomes symmetric and all eigenvalues converge onto the real axis (upper panels). Consequently, the spectrum distribution marginalized onto the real axis realizes the marchenko–pastur law (lower panels). Each histogram represents the distribution of the real parts of numerically obtained eigenvalues. Dashed lines show the theoretical predictions, and thick lines represent the marchenko–pastur law. (B) In the limit of large $\alpha$, the eigenvalue distribution of the product of correlated matrices converges to the elliptic law. As $\alpha$ increases by increasing $M$, the eigenvalues become uniformly distributed within an elliptical region and converge to the uniform distribution described by the elliptic law.}
  \label{fig_asymtotic}
\end{figure}

We first consider the limit $\tau\to 1$, where the matrix $J$ becomes symmetric because $U=V$. In this case, the elliptical region shaping the bulk density collapses onto the real axis, and integrating Eq.~\eqref{rho_supple} over $y$ recovers the Marchenko–Pastur law. We define $\alpha_+$ and $\alpha_-$ as the right and left edges of the ellipse in the limit $\tau\to 1$, which are given by $a_{\pm}=(\beta\pm1)^2/\beta$. When $x \in [a_-,\ a_+]$, we obtain

\begin{align}
    \lim_{\tau\to 1}\int\rho_b(x+iy)dy
    &= \lim_{\tau\to 1}
    \frac{\beta\left(1+\tau^2\right)}{2\pi\left(1-\tau^2\right)}\left(\log\left(1+\frac{Y}{\sqrt{k^2+x^2+Y^2}}\right) 
    -\log\left(1-\frac{Y}{\sqrt{k^2+x^2+Y^2}}\right)\right)
    \nonumber \\
    &= \lim_{\tau\to 1}\frac{\beta}{2\pi}\cdot
    \frac{\sqrt{(1+\tau^2)^2-\left(x-\tau(\beta+1/\beta)\right)^2}}
    {\sqrt{x^2+(1-\tau^2)^2
    \left\{
        1-\left(
            \frac{x-\tau\left(\beta+1/\beta \right)}{1+\tau^2}
        \right)^2
        +\frac{1}{4}\left(\beta+\frac{1}{\beta}\right)^2
    \right\}}}
    \nonumber \\
    &=\frac{\beta}{2\pi x}\sqrt{\left(a_{+}-x\right)\left(x-a_{-}\right)}.
\end{align}
Here, we used Eq.~\eqref{y} in the first line, and applied a Taylor expansion in the second line, noting that $Y$ becomes small in the limit $\tau \to 1$. Combining this with the fact that $\rho_b = 0$ for $x \notin [a_-, a_+]$, this result reproduces the Marchenko–Pastur law:
\begin{align}
  \lim_{\tau \to 1}\int \rho_{\rm{b}}(\omega)dy =
  \begin{dcases}
    \frac{\beta}{2\pi x}\sqrt{\left(a_{+}-x\right)\left(x-a_{-}\right)}
     & \text{if } a_{-}<x<a_{+}\\
     0 & \text{otherwise}
  \end{dcases}.
\end{align}
The validity of this convergence is confirmed in Figure \ref{fig_asymtotic}.

Then, we consider another limit, $\alpha \to \infty$, where $M$ becomes infinitely large compared to $N$. In this limit, by the central limit theorem, the off-diagonal elements of $J$ become Gaussian random variables with zero mean and variance $1/N$, and are independent of each other except for correlations between transposed pairs, $\langle J_{ij} J_{ji} \rangle = \tau^2/N$. The diagonal elements also converge to Gaussian variables, but with nonzero mean $\langle J_{ii} \rangle = \tau \beta$.

Noting that the ellipse is shifted by $\tau \beta$ from the origin due to the nonzero diagonal elements, the bulk spectral density $\rho_b(\omega + \tau \beta)$ is expected to converge to the elliptic law with zero mean and variance $\tau^2/N$ in this limit. Indeed, taking the limit in Eq.~\eqref{y}, we obtain
\begin{align}
    \lim_{\alpha \to \infty}\rho_{\rm{b}}\left(\omega+\tau\beta\right) &=\lim_{\alpha \to \infty}
        \frac{\beta}{2\pi(1-\tau^2)}\left(\left(\frac{1-\tau^2}{2}\left(\beta-\frac{1}{\beta}\right)\right)^2+(x+\tau\beta)^2+y^2\right)^{-1/2} \nonumber \\
    &=\frac{1}{\pi(1-\tau^4)^2}.
\end{align}
Since it is also evident that $\rho_b = 0$ outside the ellipse, we conclude that the spectrum distribution converges to the full expression of the elliptic law:
\begin{align}
  &\lim_{\alpha \to \infty}\rho_{\rm{b}}\left(\omega+\tau\beta\right) =
  \begin{dcases}
    \frac{1}{\pi(1-\tau^4)^2}
    & \text{if } \left(\frac{x}{1+\tau^2}\right)^2
    +\left(\frac{y}{1-\tau^2}\right)^2<1\\
    0 & \text{otherwise}
  \end{dcases},
\end{align}
Figure \ref{fig_asymtotic} verifies the validity of this result.

